\documentclass[11pt,twoside]{article}
\usepackage{acta-info,euler}

\newcommand{\vol}{\textbf{2,} 1 (2010) 99--112}   %% to be completed by the editor, do not modify it
\newcommand{\amss}{\amssubj{%
68N18
}}     %% Source:   http://www.ams.org/msc/
\newcommand{\CRs}{\CRsubj{%
D.1.1 
}}   %% Source: http://oldwww.acm.org/class/1998/
\newcommand{\keyw}{\keywords{functional programming, public-key
cryptography }}

\begin{document}
\title{Public-key cryptography in functional programming context}

\maketitle

\oneauthor{\href{http://www.ms.sapientia.ro/~mgyongyi}{Gy\"ongyv\'er
M\'arton}
}{%
 %% Write here your affiliation including maybe your address using \href command. You can use \\ for line breaks.
\href{http://www.sapientia.ro}{Sapientia Hungarian University
of Transylvania}\\ Cluj \\
\href{http://www.ms.sapientia.ro}{Department of Mathematics and Informatics}\\
{Tg. Mure\c s, Romania } }{
 \href{mailto:mgyongyi@ms.sapientia.ro}{mgyongyi@ms.sapientia.ro}
}
\short{Gy. M\'arton }{PKC in functional programming context }

\begin{abstract}
Cryptography is the science of information and communication
security. Up to now, for efficiency reasons cryptographic
algorithm has been written in an imperative language. But to get
acquaintance with a functional programming language a question
arises: functional programming offers some new for secure
communication or not? This article investigates this question
giving an overview on some cryptography algorithms and presents
how the RSA encryption in the functional language Clean can be
implemented and how can be measured the efficiency of a certain
application.
\end{abstract}

%% Write here your paper using the section command, theorem-like environments as:
%% definition, theorem, lemma, corollary, criterion, example, exercise, notation, problem, proposition, remark,
%% conjecture. All these are numbered consecutively.
%% For proof you can use the proof environment.

\section{Functional programming}

Functional programming is based on the lambda-calculus, with its
main principle developed in 1930, by Alonzo Church and Stephen
Cole Kleene. With lambda-calculus we can define the notion of
computable function, and using the rules defined by
lambda-calculus we can express and evaluate any computable
function. But in the early development stage of computer science
the computer calculating mechanisms were based on another
calculating model, such as a Turing machine's. The Turing machine
functioning mechanism was described in 1936 by the mathematician
Alan Turing, in the same time as lambda calculus. The Turing
machine was the base of Neumann's architecture computer model and
determined the developmental direction of programming language. In
consequence most of the real-world applications was based on
imperative programming language.

Only with the development of software technologies it became
possible to design programming languages based on stable
mathematics like functional programming language and declarative
programming language: Lisp (1958), Prolog (1970), Haskell (1990),
Clean (1995) and so on. But the usage of these languages in the
real-world applications is still very little. Some  well-known
applications of functional languages are the Yahoo shopping
engine and  the telecommunication software design by Ericsson.

A program in a functional programming language consists of a
collection of function definition and an initial expression. So
the basic method of computation is the application of functions to
arguments with the goal to evaluate the initial expression. The
basic concepts and the basic elements of a functional programming
language can be defined as follows:
\begin{itemize}\addtolength{\itemsep}{-0.5\baselineskip}
    \item persistent data structures: data once built never
    changes
    \item recursion: primary control structure
    \item high-order functions: functions that take functions as argument and in case return
function as result.
\end{itemize}
By contrary to functional programming in the imperative
programming style the main goal is to follow the changes of the
states. The basic concepts of an imperative language are:
\begin{itemize}\addtolength{\itemsep}{-0.5\baselineskip}
    \item mutable data structures: with assignment statement we can change the value
of some data
    \item looping: the primary control structure
    \item first-order programming: we cannot operate with function as
    first-class entities.
\end{itemize}
The Clean programming language, that we are use for implementing
the RSA file encryption scheme is a pure and lazy functional
language, with a fast compiler. Was developed on the University of
Nijmegen, Netherlands. The first version was released in 1995.
This language has most of the characteristic that a functional
language must to have, some of them: isn't allowed the destructive
updates; has the referential transparency property; the basic
computation form is the recursion; we can use high order
functions; it is strongly typed; list comprehension is allowed;
polymorphism is allowed too and so on.

In the real world one of the most known software for cryptography,
writing in functional programming language is Cryptol \cite{CRY}.
It was designed by Galois Connections Inc. in consultation with
expert cryptographers between 2003-2008. As developers say Cryptol
is a high-level specification language for cryptography that means
programmers who use Cryptol can focus on the cryptography itself,
and are not attended by machine-level details. In the same time
they can deal with low-level problems, namely it can work with low
level data, such as array of bits. The code written in Cryptol can
be converted to other languages such as Haskell, VHDL and C. It
can use with various platform such as embedded systems, smart
cards and FPGAs.

Another well-known software package in this field is the
RSA-Haskell, \cite{SAN}, which was published in 2007, by David
Sankel. It is written in Haskell and as the author says it is a
"collection of command-line cryptography tools and a cryptography
library". With RSA-Haskell using command-line tools users can do
secure communication: encrypt/decrypt some message, can identify
the sender and authenticate the message. The crypto library is
licensed under the GPL (General Public License), and allowed for
users to access some cryptography algorithms to incorporate these
in their application. The size of the RSA-key is 2048 bit, and
SHA512 hash algorithm is used in conjunction with OAEP (Optimal
Asymmetric Encryption Padding), \cite{TIL}.

\section{Cryptography}
In the past Cryptography was the science of secret codes. Today
due to the electronic world the data security is in the center of
the attention. This means that it became more than producing
secrete code. The topmost tasks are privacy, integrity,
authentication, and nonrepudiation.

Cryptography algorithms can be classified in two groups such as
symmetric-key cryptography and public-key cryptography according
to what kind of keys are in use in the system: secret or public
keys. In the case of symmetric-key cryptography the processes of
encrypting and decrypting are coming to pass with the same secret
key, in contrast to public-key cryptography where the encrypting
are coming to pass with public key, and the decrypting with an
other key, with private key. Besides encryption schemes more other
schemes belong to the public-key cryptography such as digital
signatures schemes, authentications schemes, and so on \cite{SAL}.
But on backwards our attention will focus only on public-key
encryption schemes. So, now we will give the formal definition of the public-key
encryption scheme, \cite{HOF}:

\begin{definition}
A public-key encryption scheme with message space $\mathbb{M}$ can
be define with three algorithms: $PKC = (GEN, ENC, DEC)$, where
\begin{itemize}\addtolength{\itemsep}{-0.5\baselineskip}
    \item $GEN$ is the key generation algorithm, which determines in a
random way the public and secret key-pairs: $(p_k, s_k) =
GEN(\epsilon)$, where $p_k$ is the public key and $s_k$ is the
secret key,
    \item $ENC$ is the encryption algorithm, which encrypts a
    message $M$, producing the ciphertext: $C =
    ENC_{p_k}(M)$, where $M \in \mathbb{M}$,
    \item $DEC$ is the decryption algorithm, which decrypts the
    ciphertext $C$, $M = DEC_{s_k}(C)$.
\end{itemize}
\end{definition}
For the correctness of the system we require that
$DEC_{s_k}(ENC_{p_k}(M)) = M$. For the security of the system the
corresponding requirements can not be claim so easy. One of them
is that the public key inversion problem (finding the secret key
for a given public key) must be based on a hard mathematical
problem, in average case of the problem instances. Another
requirement is that the ciphertext inversion problem (finding the
encrypted $M$ message for a given $C$ ciphertext and $p_k$ public
key) must be hard. But very few problems we are known that
achieves these requirements, thus not surprising that in a real
word application the most of the cryptography systems are based on
the following mathematical problems:
\begin{itemize}\addtolength{\itemsep}{-0.5\baselineskip}
    \item factoring large integers, for
instance the RSA cryptosystem
    \item computing discrete logarithms, for
instance the ElGamal cryptosystem.
\end{itemize}
For these problems no one knows polynomial time algorithms,
moreover these problems are those scarce problems that are not
classify between the P and NP-complete classes, \cite{SAL}.

Because in our implementation we concern on RSA file encryption
scheme, first we will present the basic RSA encryption scheme,
\cite{RSA}. This scheme consists in three steps, corresponding to
the 3 algorithms specified in the formal definition of public-key
encryption scheme: key generation, encryption, decryption. The
message space is the $\mathbb{M} = \mathbb{Z}_m^*$, where $m$ is
an integer number, calculated in the key generation step. The
value of $m$ determines the order of magnitude of the RSA-key.
\begin{itemize}\addtolength{\itemsep}{-0.5\baselineskip}
    \item Generating the RSA keys consists on the following steps, where
$\phi$ is the Euler function, and $gcd$ is the greatest common
divisor of the arguments, \cite{MAR}:
\begin{itemize}\addtolength{\itemsep}{-0.5\baselineskip}
    \item generating two big random prime numbers: $p, q$,
    \item calculates the product $m = p \cdot q$, henceforth $m$ we will call
    modulus,
    \item selects randomly $e$, where $1 \le e \le \phi(m)$ and
$gcd\,(e,\phi(m))=1$, henceforth $e$ we will call encryption
exponent,
    \item computes $d$, where: $1
\le d \le \phi(m)$, $d \cdot e = 1\pmod{\phi(m)}$, henceforth $d$
we will call decryption exponent,
    \item the public key consist: $(e, m )$,
    \item the private key consist: $( d, m )$.
\end{itemize}
    \item For encryption of $M\in \mathbb{Z}_m^*$ we do: $C = M^e
\pmod{m}$.
    \item For decryption of $C\in \mathbb{Z}_m^*$ we do: $M = C^d
\pmod{m}$.
\end{itemize}

\section{Algorithms in Clean}
Now we shall present the implementation details of the RSA file
encryption.

Firstly we mention that arithmetic with large numbers is quite
easy in Clean, through importing the BigInt library, \cite{CLE}.
Henceforth we give the definition of this importing as well as the
definitions of constants, are used several times in the system to
be realized.

\begin{verbatim}
import BigInt
my_one :== toBigInt 1
my_two :== toBigInt 2
my_zero:== toBigInt 0
alph :== toBigInt 256
\end{verbatim}

By the way of implementing RSA file encryption system, several
questions are coming up: to find the modular multiplicative
inverse; to perform the modular exponentiation; to generate big
(more than 100 digit) random prime number; to convert the number
from the base $p$ in the base $p^k$ and inverse; to perform the
RSA encryption on numbers; do the file I/O task; create a
graphical interface.

In the following sections, one after another we will briefly
present how we are resolved this questions in the Clean
programming language.
\subsection{Multiplicative inverse}
In order to find the multiplicative inverse of an integer $a$
$\pmod{b}$ we need to resolve the congruence: $a \cdot x_1 =
1\pmod{b}$ with the unknown coefficient $x_1$. This congruence can
be solved by using the extended Euclid's algorithm, \cite{BEG}. For
this we write two functions:
\begin{itemize}\addtolength{\itemsep}{-0.5\baselineskip}
    \item an auxiliary function: $seuclid$, with the role of doing the
proper computation, namely to calculate the coefficient of a,
    \item the main function
$meuclid$ doing the necessary initialization and the first call of
$seuclid$.
\end{itemize}
The Clean code that do this is the following:
\begin{verbatim}
seuclid :: BigInt BigInt BigInt BigInt->BigInt
seuclid a b x1 x2
    |b==my_zero = x2
    |otherwise = seuclid b (a rem b) (x2-(a/b)*x1) x1

meuclid :: BigInt BigInt -> BigInt
meuclid a b
    #res = seuclid a b my_zero my_one
    |res<my_zero = (res+b) rem b
    |otherwise = res rem b
\end{verbatim}

\subsection{Modular exponentiation}
The modular exponentiation calculates the value of $x^p\pmod{m}$
using the fast exponentiation technique, \cite{ROS}. For this
purpose we write the function, $mexp$.

The Clean code that do this is the following:
\begin{verbatim}
mexp :: BigInt BigInt BigInt-> BigInt
mexp x n m
    |n == my_zero = my_one
    #x1 = (x*x) rem m
    |isOdd n = (x * mexp x1 (n/my_two) m) rem m
    = mexp x1 (n/my_two) m
\end{verbatim}

\subsection{Random prime number generation}
To set the RSA-key we need prime random numbers. For testing if a
random number with the right size is prime or not we use
probabilistic primality test which quickly eliminates the
composite numbers. For this test we use the Miller-Rabin primality
test \cite{ROS}.

The formal definition of this test is the following:
\begin{definition}
Let the odd number $q$ and $j$ such that $n-1 = 2^{j}q$. If the odd
number $n$ is prime, then for all numbers $x$, where $gcd\,(x,n) =
1$ one of the following statements is true:
    \begin{itemize}\addtolength{\itemsep}{-0.5\baselineskip}
        \item $x^{q}= 1\pmod{n}$,
        \item for one of the $i$: $x^{2^{i}q}=n-1\pmod{n}$, where $0 \le i \le j-1$.
    \end{itemize}
\end{definition}

For that in the Clean programming language we write the function
$mfind$ which determines the value $j$ and $q$ in the exponent
$n-1$ by calling the $sfind$ function. To test the second
statement of the above definition we write the $subtest$ function.
The validity of the first statement is tested by the main function
$millerrabin$. For this purpose we calculate the value
$x^{q}\pmod{n}$ with the modular exponentiation algorithm.

The Clean code that do this is the following:
\begin{verbatim}
millerrabin :: BigInt BigInt -> Bool
millerrabin n x
    #(q, j) = mfind n
    #y = mexp x q n
    |y == my_one = True
    = subtest n y j

mfind :: BigInt -> (BigInt, BigInt)
mfind n = sfind (n - my_one)my_zero

sfind :: BigInt BigInt -> (BigInt, BigInt)
sfind q k
    |isEven q = sfind (q / my_two) (k + my_one)
    = (q, k)

subtest :: BigInt BigInt BigInt -> Bool
subtest n y j
    |j < my_one = False
    |y == (n - my_one) = True
    |y == my_one = False
    = subtest n ((y * y) rem n) (j - my_one)
\end{verbatim}

For generating random number we use the algorithm of linear
congruential generator \cite{ROS}.

To be certain of that an odd number $n$ is prime, we must have the
result 'True' in the $millerrabin$ function, for more, different
$x$ random numbers. For that we examine the number $n$ if it is
odd or it is even, and in the case it is odd we generate $t$ bit
random number and test $t$ times for these $x$, if the
$millerarbin$ function is 'True' or not.

\subsection{Base expansion}
In the RSA encryption scheme we can encrypt only a single large
number. In RSA file encryption we must to encrypt many bytes. To
achieve this we do a pre-processing on bytes that we want to
encrypt. We choose $k$ bytes, that corresponds to the size of the
block that we are encrypt all at once. We see these bytes as
digits in base 256 and we make a conversion from 256 to base
$256^k$ \cite{MAR}. After that we have a single big number which
we encrypt corresponding the scheme presented in section 2. The
Clean function which resolves this conversion is $myconvert1$ by
the help of auxiliary function $subconvert1$.

The Clean code that do this is the following:
\begin{verbatim}
myconvert1 :: BigInt BigInt BigInt-> [Char]
myconvert1 nr p pk = subconvert1 nr p pk my_one []

subconvert1 :: BigInt BigInt BigInt BigInt [Char] -> [Char]
subconvert1 nr p pk nr1 tomb
    | nr1 >= pk = tomb
    | otherwise = [(toChar (toInt (nr rem p))) :
                    (subconvert1 (nr/p) p pk (nr1+my_one) tomb )]
\end{verbatim}

After encryption in the process of decryption we must do the
inverse conversion, so we need an algorithm that makes the
conversion from $p^k$ to $p$.

The Clean code that do this is the following:
\begin{verbatim}
myconvert :: [Char] BigInt BigInt -> BigInt
myconvert n p pk = subconvert n p pk my_zero my_one

subconvert :: [Char] BigInt BigInt BigInt BigInt -> BigInt
subconvert [] p pk nr exp  = my_zero
subconvert [kezd : veg] p pk nr exp
    | pk == nr = my_zero
     = (toBigInt (toInt kezd)) * exp +
            (subconvert veg p pk (nr+my_one) (exp*p) )
\end{verbatim}

\subsection{RSA encryption on numbers}
To attain the RSA encryption first we generate the private, secret
key-pairs, after that we perform encryption/decryption with this
keys. For efficiency reason, but without loss of security we have
choose $e$ as a constant: $65537$ to be encryption exponent in the
public key pair. This choice is commonly used in practice to speed
up encryption. In contrast, for security issues, to avoid the
small decryption exponent attack, decryption exponent can not be
too small. $d$ must have approximately the same size as modulus
$m$, \cite{MOV}. These choices determine the encryption and
decryption time, so decryption time always will be much longer
than encryption time. Several technics were developed to shorten 
the decryption time, one of them is using the Chinese remainder
theorem. But even with these technics RSA encryption/decryption is
much slower than the commonly used symmetric-key encryptions
methods.

After that we generate two big prime numbers and calculate their product $m$. 
Now using the multiplicative inverse function we
can calculate the private key $d$ as an inverse of integer $e$
modulo $\pmod{\phi(m)}$.

The Clean code that do this is the following:
\begin{verbatim}
privatk :: BigInt BigInt BigInt -> BigInt
privatk e p q
    #pq = (p - my_one)*(q - my_one)
    |gcd e pq <> my_one = abort "not relative prime"
    = meuclid e pq
\end{verbatim}

To encrypt a number $x$, where $1 < x < ( m-1 )$ and $gcd\,(x, m)
= 1$ we can do one modular exponentiation: $c = x^e\pmod {m}$. So
the magnitude of the modulus $m$ determines the magnitude of the
number that we can encrypt at one go, and in the same time
basically determines the running time of the application.

The Clean code that do this is the following:
\begin{verbatim}
rsacrypt :: BigInt BigInt BigInt -> BigInt
rsacrypt x e m = mexp x e m
\end{verbatim}

For decryption we do the same computation, the differences consist
in the value of actual parameter, we use the decryption exponent
$d$ instead of $e$.

\subsection{File I/O}
Because the Clean is a pure functional language the destructive
updates are not admissible, but when we dealt with a file I/O we
must to have the possibility to destructively update the file. In
Clean this situation was resolved by introducing a new type, for
which we use the * notation in the type name. With this type we
can restrict the references of some date structure, and when the
reference is unique, the update of date structure is allowed
\cite{CLE}. Using this technics the file I/O is quite simple in
Clean.

To obtain more security in file encryption, several block cipher
technics can be use. The most common block ciphers are the ECB
(electronic codebook), CBC (cipher-block chaining), CFB (cipher
feedback) and OFB (output feedback) modes. In our implementation
we use the ECB mode, and we use the same key for every block
encryption/decryption.

As long as we want to use the RSA file encryption as a block
cipher we must set $k$, the size of the blocks. We assume that the
plain text is a binary file so the alphabet size that we are using
is 256, corresponding to possible byte values, thus the size of a
block is $log_{256}m$, where $m$ is the RSA modulus.

Now we present the function $filecrypt$ which has the role:
testing if we are at the end of the plain text file; reading $k-1$
bytes from this file; calculating $k2$, the number of bytes that
were effectively read; converting these bytes from base $p^k$ to
base $p$; encrypting the result; converting the result to base
$p^{k+1}$; writing in an encrypted file the number $k2$; writing
the $k+1$ bytes in an output file, that will be the encrypted
file.

The Clean code that do this is the following:
\begin{verbatim}
filecrypt :: *File *File (BigInt, BigInt) -> (*File, *File)
filecrypt inf outf (e, n)
    #! (atEnd, inf) = fend inf
    |atEnd = (inf, outf)
    # k = nrblok n alph
    # (inf, res) = mread (k-my_one) inf
    # (k2, cr) = filecrypt` res k e n
    # outf = fwritec (toChar k2) outf
    # outf = mwrite cr outf
    = filecrypt inf outf (e, n)

filecrypt` :: [Char] BigInt BigInt BigInt -> (Int, [Char])
filecrypt` res k e n
    #k2 = length res
    #nr = myconvert res alph (toBigInt k2)
    #scr = rsacrypt nr e n
    #cr = myconvert1 scr alph (k+my_one)
    = (k2, cr)
\end{verbatim}

The function $mread$ used in above code has the role to read a
given number of bytes from a file while the function $mwrite$ has
the role to write a list of bytes in a file.

\begin{figure}[t]
\begin{center}
\includegraphics[scale=0.8]{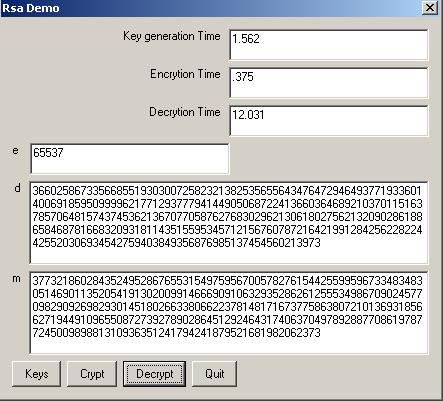}
\end{center}
\caption{The graphical interface}
\end{figure}

\subsection{The graphical interface}
In Clean using the Object I/O library we can write flexible,
platform independent programs, with a well designed graphical user
interface, \cite{AWI}. For obtaining an easy method to manipulate
our input/output, such as generating public/secret key; showing
the key generation time; selecting input/output file; doing the
encryption/decryption; showing the encryption/decrytion time we
design a graphical interface with different dialog items such as
button controls, edit controls, text controls.

Our application graphical interface, for a certain jpg files with
48.9 KB size, and for 1024 bit key size have the following
appearance, where we made out the value of public, secret keys (do
this only for testing case), the key generation time for these
values in seconds, and the encryption/decryption times in seconds are given in Fig. 1. 

To measure the running time of certain function we got the
computer tick, and determined the difference between two ticks.
For this we have the following built in functions:

\begin{verbatim}
getCurrentTick :: !*env -> (!Tick, !*env)
tickDifference :: !Tick !Tick -> Int
\end{verbatim}

To measure the performance of an application we have another
possibility that Clean offers: we can enable Time Profiling option
insight in the Clean environment, which means that after the
execution the program will write a profile file. For our
application the generated profile file is given in Fig. 2.

\begin{figure}[t]
\begin{center}
\includegraphics[scale=0.75]{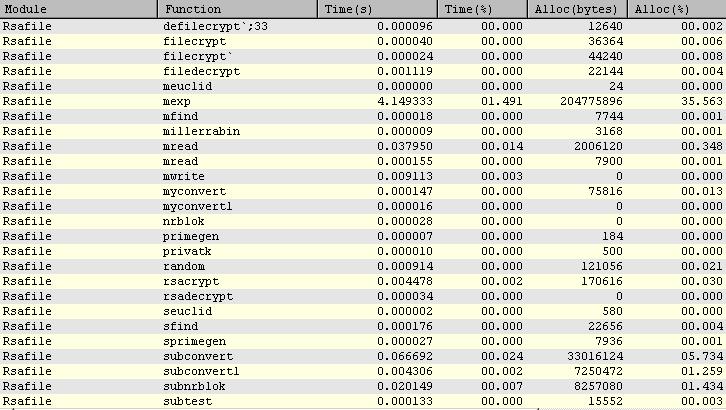}
\end{center}
\caption{The content of profile file}
\end{figure}

This profile file is overwritten after every program execution and
will consist the real time measurements of each function in
seconds, the number of bytes allocated in the heap by the function
an so on.

\section{Conclusions}
Most of cryptography applications assure security on hardware and
software level too. The main disadvantage using functional
programming in cryptography is that applications written in
functional programming language can guarantee security principally
on software level. This fact constrains the applicability of
functional programming language in area of cryptography. Another
reason withdrawal in usage of functional programming in range of
security is that the community of programmers who use functional
programming language for building their cryptography's software is
relatively small, and the available documentation is still very
scarce.

Our program vulnerability is the usage of linear congruential
generator for generating random number, which is known as not very
safe. But our purpose not was to study the complex area of
pseudorandom number generator algorithms.

Our experience shows that cryptography algorithms coded in a
functional programming language are much shorter than those coded
in C, Java, Maple. It is relatively easy to test the functional
programming function independently from the entire program. The
syntax "forces" the programmer to write more modular codes, so it
is simple to locate and correct errors in these modules. Type
errors are much easier to prevent and in case, to correct. We can
easy issue efficiency in time and space. The process of the file
I/O are relatively simple. It has a built-in library for large
numbers, which means that working with large numbers becomes quite
simple. The concept and syntax are permitting the correct use of
necessary mathematics, so the programmer can focus on these, and
not loosing time on circumstance of implementations. For those who
know functional language it is easy to read and understand Clean
code.

Taking account of our algorithms time consuming and correctness,
and considered the capacity of software package presented in the
first section we can establish that the usage of functional
programming language can't be considered inconvenient. So we can
conclude that functional programming is a very useful tool to
write stable an efficient cryptography applications.

\bigskip
\rightline{\emph{Received: December 5, 2009  {\tiny \raisebox{2pt}{$\bullet$\!}} Revised: February 23, 2010}}     %% to be completed by the editor

\end{document}